# Disturbance Ratio for Optimal Multi-Event Classification in Power Distribution Networks

M. D. Borrás, J. C. Bravo, and J. C. Montaño, *Senior Member, IEEE*

*Abstract*— This paper presents an effective approach to identify power quality events based on IEEE Std 1159-2009 caused by intermittent power sources like those of renewable energy. An efficient characterization of these disturbances is granted by the use of two useful wavelet based indices. For this purpose, a wavelet-based Global Disturbance Ratio index (GDR), defined through its instantaneous precursor (Instantaneous Transient Disturbance index ITD(t)), is used in power distribution networks (PDN) under steady-state and/or transient conditions. An intelligent disturbance classification is done using a Support Vector Machine (SVM) with a minimum input vector based on the GDR index. The effectiveness of the proposed technique is validated using a real-time experimental system with single events and multi-events signals.

*Index Terms*—IEEE Std 1159-2009, Power quality (PQ), single-event and multi-event power quality, SVM, wavelet transform.

## I. INTRODUCTION

NOWADAYS the increasing feeder development and consumption of renewable energy as well as the extensive use of power-switching devices for source conditioning and motion control in modern industrial applications has increased the eventuality of unacceptable harmonic levels, poor power factor, and unbalanced currents and voltages in power distribution networks (PDN) [1]. In this power quality (PQ) panorama, the effective detection and identification of disturbances become the most important task for the protection and selection of effective mitigation techniques in PDN.

In the area of PQ analysis, some advanced mathematical techniques such as short-time Fourier transform (STFT)[2]-[3], S-transform [4]-[6], wavelet transform (WT)[7]-[10], Hilbert-Huang Transform[11], Kalman filter[12], strong trace filter (STF) [13], sparse signal decomposition (SSD)[14], Gabor-Wigner transform [15] and empirical mode decomposition (EMD)[16]-[17] were used to extract the feature eigenvectors that enabled the disturbance identification. Despite these tools allowed a classification stage too, most of them were restricted mainly to single disturbances. To include multi-event signals improving the classification process, some artificial intelligence techniques such as artificial neural network (ANN)[18], support vector machines (SVM)[19]-[23], fuzzy expert systems and genetic algorithms (GA) were employed[12],[24].

The neural tree based method presented in [17] classifies eight disturbances with a high accuracy and the work proposed in [18] uses neural networks to identify single and combined PQ events. However both of them require a high computational burden which causes relevant time delay in classification, and spend a lot of time in the training stage. On the other hand, the SVM based methods offer a great potential to handle large features, provide stable solutions to quadratic optimization with high learning processes, but they present poor classification accuracy when training samples are minimum. In this way, the main advantage in [16] is the correlation preservation between different event types which improves the accuracy of the method. Nevertheless, the data set is high to include all characteristic of multi-event signals. Additionally, due to the complex classification scheme of Rank Wavelet SVM, signals may be identified to multiple classes. The algorithm presented in [23] uses a set of simple binary SVM classifier but requests 31 parameters to characterize a disturbed signal.

Most of the aforementioned analysis needs high processing speed to manage with large computational data within a smaller time. In paper [25] a stand-alone general purpose microcontroller is used to implement a classifier based on SVM but it needs a discriminant analysis to find out the smaller set of uncorrelated features from higher order statistical data.

The work presented in this paper belongs to the actual trending focused on enabling the performance of sensing devices using fast and low-cost hardware able to be easily improved for real-life applications.

The number of features required in the proposed approach is less than any of the precedent works cited above. It is due mainly to the proper choice of two selected PQ indices. A power quality index (PQI) is the summarization of waveform distortions in electrical signals from the perfect sinusoids. That is the comprehensive but expressive presentation of the impacts of the non-ideal waveforms to electric power systems and is extendable to accommodate wider application conditions. PQIs are employed to characterize the degree of the quality degradation in a quantitative manner. Other existing indices

M. D. Borrás and J. C. Bravo are with the Electrical Engineering Department. University of Seville, 41011 Seville, Spain (e-mail: borras@us.es, carlos_bravo@us.es).
J. C. Montaño is with the Electronic Technology Department, University of Seville, 41011 Seville, Spain (e-mail: jcmontano@us.es).

such as those of Total Harmonic Distortion (THD), Power factor, Flicker factor and so on, reflect the degree of power disturbance in each of these categories individually, but fail to assess most of the above mentioned phenomena together in a comprehensive and concise way by a single value [26]-[34].

In this paper a recent wavelet-based power quality index, designated Global Disturbance Ratio (GDR) and obtained from its time-varying precursor named Instantaneous Transient Disturbance index (ITD), is used. They integrally assess the power transfer quality in steady-state and/or transient situations. The used ITD index shows instantly the PQ evolution. This has the advantage of evaluating the PQ in real-time situations under whatever conditions, for any load in a PDN and extracting the disturbance characteristics. Moreover, the GDR has the benefit of assessing the PQ by means of a single value, and permits to distinguish between different events so the GDR can be used as a classifier input. They can also be further used to evaluate both the effectiveness and dynamic responses of PQ mitigation equipment in practical applications [42].

Measurement of the instantaneous power-system frequency [35]-[36] has been proposed to prevent the discrete Fourier transform (DFT) problems [37] and overcome the non-ideal dynamic response speed of discrete wavelet transform (DWT) [38]. Using a technique like this, it is possible to estimate fundamental voltage and current, their wavelet components as well as derived electrical parameters accurately with satisfactory dynamic, and free of system frequency fluctuation, even under asynchronous sampling.

In this work, the GDR index and RMS value are used directly as inputs of a SVM based on the *one versus one* (OVO) multiclass method to classify disturbed signals of real power systems. It performs a fast, easy and low computational cost SVM classifier by using single-event as well as multi-event signals at the training and test stages. So, stationary and/or nonstationary disturbance effects are identified and quantified by means of this procedure. Moreover, feature data size is reduced to a simply two dimensional vector representing the characteristics of any disturbed electric signal. Real-time results have confirmed the high resolution and accuracy of this technique.

Understanding of signal processing considerations, input vector based on PQ indices and classification using SVM are considered in Section II. The measurement process based on a specific developed system and the obtained results are presented in Section III.

## II. Theoretical Concerns

### A. Signal Processing Considerations

*1) Preprocessing stage: Frequency variation*

Fast and accurate system-frequency measurement is fundamental requisite to obtain with exactitude root-mean-square (RMS), power and energy values. Under system frequency variation, the number of samples per period of the original signal do not equal an integral value, which must be a power of two in FFT and DWT calculation. Thus, an Instantaneous Frequency Calculation (IFC) is achieved as a first step of the GDR calculation to avoid errors due to lack of synchronization between the signal period and the sampling sequence.

The IFC is based on the frequency estimation of the signal using three equidistant samples for a nominal frequency of 50 Hz and an observation window of 10 cycles, corresponding to the IEC61000-4-30 standard window [45]. The corresponding algorithm diminishes the variance of the estimation [36]. By this procedure, single and three-phase networks are treated, and uncertainty in the frequency estimation under severe conditions of signal quality is obtained.

Hence, samples of signal $s(n)$ are processed,

$$s(n) = A \cdot \sin(\omega \cdot n \cdot T_S + \varphi) \quad (1)$$

where $\omega = 2\pi f$; $T_S$ is the sampling period, $n$ the samples number, and $\varphi$ the phase angle.

For single-phase systems, this expression can be evaluated from equidistant samples [.., $s_{-1}$, $s_0$, $s_1$,..]. In the case of three samples, the following expression can be obtained [36]:

$$c = \frac{\sum_{j=0}^{2}(s'_0)_j \cdot \left[(s'_{-1})_j + (s'_1)_j\right]}{2 \cdot \sum_{j=0}^{2}(s'_0)_j^2} \quad (2)$$

where $c = cos(\omega T_S)$ and, therefore, $f = cos^{-1}(c)/(2\pi T_S)$. Samples $s'_n$ are supposed random variables, that is, they are the sum of true samples $s_n$ containing the theoretic value [36] and noise, which shows normal distribution, with zero mean and standard deviation.

The error occurs, in general, when the interval between samples, $T_S$, is different from $2\pi/N$, where $N$ is the number of data points per period. A process to adjust this interval, in all circumstances, is therefore necessary. In this case, to correct DWT errors, the constant sampling number in one period of the input signal is kept, and the sampling frequency according to the frequency of the input signal is modified by using the IFC information. Then, corrected samples are obtained to improve further digital processing based on the DWT and accurate RMS values.

*2) Feature extraction stage*

Due to their powerful features, the DWT and multiresolution analysis (WMRA) are chosen in this work for the joint analysis of the stationary and transient parts of electrical signals. These parts are extracted from a monitoring window to assure the correct use of the DWT.

However, the DWT uncertain limitation may be a lack that can be minimized by an appropriate choice of the analyzing wavelet basic function [39]. In this paper, the Meyer wavelet offers the best results for the analyzed electrical signals present in PDN.

The aim of WMRA is to develop representations of a complicated signal in terms of scaling and wavelet functions. The amount of WMRA decomposition levels is limited by the number of samples of the signal, which in turn must be a power of two.

Thus, the signal $s(n)$ can be presented in terms of its frequency components, i.e., coefficients $a_{J,k}$, $k = 1,..,2^J$, representing the smoothed version of signal $s(n)$, and coefficients $d_{j,k}$, $k = 1,..,2^J$, $j = 1,\ldots,J$ characterizing the detailed versions of $s(n)$. They contain the lower and higher-frequency components respectively [9]. So, the complete signal can be expressed as

$$s(n) = a_J(n) + \sum_{j=1}^{J} d_j(n) \qquad (3)$$

For extracting the fundamental component of a waveform by using WMRA, the sampling rate of the signal and, so, the number of WMRA steps, $J$, must be specified.

In this work has been assumed that the most important transients occurring in actual situations of power systems are captured into a frequency band of 6400 Hz, for $fs = 1/T_s = 12.8\ kHz$. It assures accurate results with $J = 6$.

*B. Input indices vector*

Two parameters, named $k_1$ and $k_2$, are used as inputs in the classification state. This pair of feature parameters is made up of the RMS value and the GDR index.

*1) DWT-based RMS calculation*

Through WMRA decomposition (3), the RMS value of signal $s(n)$, $S$, can be obtained

$$S = \sqrt{A_J^2 + \sum_{j \leq J} D_j^2} \qquad (4)$$

where $A_J$, is the rms value of the $N$ samples signal $a_J(n)$ in the lowest frequency band $J$, in which the fundamental component of the signal is included. $\{D_j\}$ is the set of rms values of $d_j(n)$ signal in the higher frequency band, or wavelet-level lower than or equal to the scaling level $J$.

*2) DWT-based GDR calculation*

The nonstationary events duration is a very relevant factor to be considered. It can be measured with high precision by the wavelet procedure used in this work. The GDR index is given by,

$$GDR = \left(1 + \frac{T_0}{T}\right) \frac{\frac{1}{N}\sum_{n=1}^{N}\sqrt{\sum_{j \leq J} d_j^2(n)}}{A_J} \cdot 100 \qquad (5)$$

where $T_0$ is the event duration of the transient disturbance, $T$ is the time interval window used and $A_J$ is the fundamental energy component defined as:

$$A_J^2 = \frac{1}{N}\sum_{n=1}^{N} a_J^2(n) \qquad (6)$$

Equally,

$$GDR = \left(1 + \frac{T_0}{T}\right)\frac{1}{N}\sum_{n=1}^{N} ITD(n) \qquad (7)$$

where the ratio ITD(n), is given in terms of the time-scale distribution of the WMRA components:

$$ITD(n) = \frac{\sqrt{\sum_{j \leq J} d_j^2(n)}}{A_J} \cdot 100 \qquad (8)$$

The selection of the time interval $T_0$ can be determined by the time index of the first maximum peak value of the ITD(n), $t_0$, and the time index of the last maximum peak value of the ITD(n), $t_0 + T_0$.

The ITD(n) can be interpreted as a "time-varying" power quality evaluation determined by the time-frequency localized energy ratio of the disturbance to the fundamental frequency energy.

Also,

$$GDR = \left(1 + \frac{T_0}{T}\right)\langle ITD \rangle \qquad (9)$$

where $\langle ITD \rangle$, is a "*transient-interval average*" of the ITD(n), over a sample interval $N$ as follows:

$$\langle ITD \rangle = \frac{1}{N}\sum_{n=1}^{N} ITD(n) \qquad (10)$$

So the GDR index permits to quantify by a single number the time-varying signature as in the case of steady-state disturbances.

A loaded power network with sinusoidal voltages return an ideal null GDR. On the other hand, a high value of GDR would indicate a high level of steady state and/or transient disturbances, with the contribution of each event aspect well defined and measured. Note that the used index GDR presents the advantage over the traditional THD of distinguishing both, transients and stationary events. It permits to give an important role to the disturbance duration.

*C. Classification Using SVM*

In this work, an optimal feature selection has been more important than the particular kind of classifier chosen, because often, a classification based on poor features almost never is saved by a good classifier. The proposed PQ classifier utilizes SVM to identify single and multiple-combined disturbances. SVM is a supervised learning tool applied for pattern recognition and classification. As the theory of SVM is well-known and widely used, the details are obviated in this section. The specifics of SVM can be found in [40]-[41].

Particularly, an OVO approach based SVM is used to process the multiple classifications of PQ disturbances. This method needs to build $k(k-1)/2$ classifiers where each one is trained using data only from two classes of the $k$ possible and a pair wise competition between all of them is performed. The SVM classifies the feature vector obtained $V_n(k_1,k_2)$ into the class that yields the final winner.

III. DESIGN OF THE EXPERIMENTAL SYSTEM AND RESULTS

The proposed classification plan of disturbance is shown in Fig. 1, and the main steps described in previous sections have been marked in dot line. The first one is the preprocessing stage, and then a fast DWT-based component calculation step is followed allowing the extraction of the feature vectors. The last stage consists in classification involving determination of PQ multi-event by using SVM.

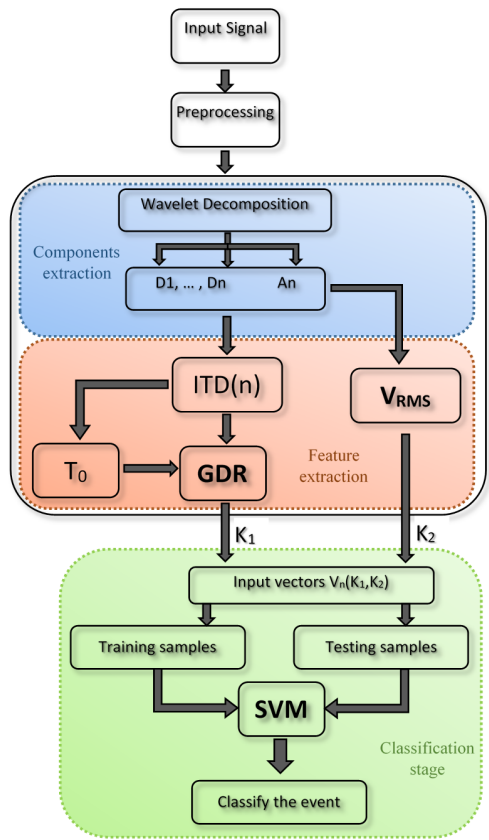

Fig. 1. Flowchart of the proposed methodology.

Next, the measurement process, the experimental test and results are detailed.

*A. Measurement Process*

Fig 2 illustrates the complete bench; it is the experimental setup used to test the effectiveness of the proposed method under common real time working conditions. This procedure permits the emulation of actual power systems and can be modeled by hardware and software components.

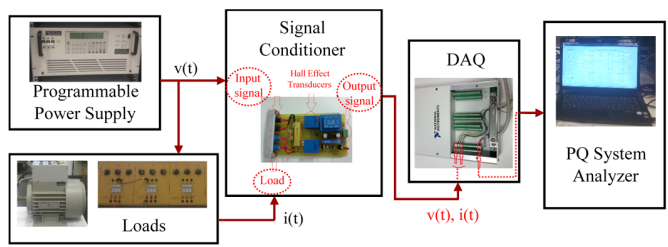

Fig. 2. Laboratory setup for real time experiments.

The hardware block contains a programmable AC power supply (Pacific Power Source 320-AMX), a specific-designed signal conditioner, a NI USB-6259 DAQ, typical loads and a personal computer (PC). The power supply is programmed to generate single or three-phase voltage patterns based on the parameters defined by the user; thus, steady-state and/or transient-state disturbances are modeled following the IEEE standard 1159-09 for monitoring electric power quality [46]. The host PC is equipped with the NI USB-6259; it is a High-Speed M Series Multifunction DAQ for USB 16 bits. It acquires eight differential inputs. Analog inputs are converted with 16 bits of resolution sampled at 1.25 MS/s. Voltage and current sensors are built with Hall Effect voltage and current transducers, type LV25-P and LA25-NP respectively. Low-level voltage signals proportional to the phase-neutral voltages are available. The power supply used fulfills all the proposed requirements:

1) output voltage up to ±600 peak volts;
2) maximum output power: 1.2 kVA;
3) bandwidth (30–5 kHz) at full power;
4) THD < 0.2%.

Disturbed voltage signals at grid level are generated for studying PQ events in several types of loads and the currents are available for evaluating the behavior from the consumer point of view.

The generated voltage signals are used to simulate a power system with typical voltage sources and arbitrary loads. This method permits to process polyphase sinusoidal voltages added with simple or multiple disturbances.

Matlab software has been used to build a Graphical User Interface that acts as a virtual device that processes the signal data file from the A/D converted connected to the signal conditioner (Fig. 2). This control program diagnoses quality aspects of the input signals, such as frequency stability, distortion level, symmetry of three-phase signals (balance between phases R, S and T), and others that can be inferred from the graphical user interface of Fig. 3.

In order to carry out this diagnosis, the system can measure and present/display the graphs (with its time evolution):

- Instantaneous frequency of the network, following its changes at intervals of measurement of one cycle, considering signals deformed and with noise.
- Harmonics, represented in phasor form using two bar charts, one for magnitudes and another for phases.
- Instantaneous PQI and coefficients of power quality indices (percent), with presentation of the data corresponding to the signals.
- DWT coefficients with signal representation of the wavelet level selected by the user.
- SVM-based disturbance classifier.

Furthermore, the PQ System analyzer computes power quantities in wavelet and Fourier domain specified by IEEE standard 1459-2010 [43].

The proposed structure can be further used in real-time for both, monitoring and detecting faults in power networks [48] and electrical machinery [49]. By means of a signal-based fault diagnosis method, the used indices can be applied to loads such as induction motors, power converters and mechanical components.

### B. Experimental test and Results

The fundamental component of the grid voltage presented in all the used test signals is distorted by harmonics and/or nonstationary events, in accordance with both, the IEEE standard 1159-2009 and the European standard EN-50160 [46]-[47].

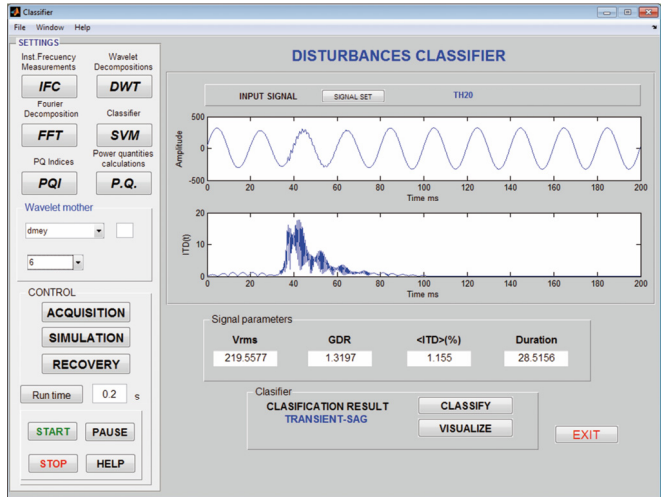

Fig. 3 Graphical User Interface of the PQ System Analyzer.

For the signal quality assessment, instantaneous frequency measurement is performed first, which enables synchronization between the signal period and the sampling sequence. For the considered voltage and current windows specified by IEC standard 61000-4-30 [45], time-frequency-based quality aspects are calculated by the DWT. For the case of the selected 12.8-kHz sampling rate, Table I gives the frequency band information for different levels of the wavelet analysis. In this paper, discrete Meyer wavelet (dmey) is chosen as the wavelet of Mallat decomposition and reconstruction for 6 layers MRA of the voltage and current signals [50].

TABLE I
FREQUENCY BANDS AND HARMONICS OF SIX LEVELS OF THE DWT

| Level | Freq. Band(Hz) | Odd Band Harmonics |
|---|---|---|
| 7 (d1) | 3200-6400 | 63rd-127rd (odd num.) |
| 6 (d2) | 1600-3200 | 33rd-63rd (odd num.) |
| 5 (d3) | 800-1600 | 17th–31st (odd num.) |
| 4 (d4) | 400-800 | 9th, 11th, 13th, 15th |
| 3 (d5) | 200-400 | 5th, 7th |
| 2 (d6) | 100-200 | 3rd |
| 1 (a6) | DC-100 | 1st |

According to equations (4), (5) and (10) the PQIs are computed. The PQ indices have been obtained from a variety of single-phase signals that includes the training signals set. In all of them, the grid voltage waveform contains a fundamental component of 230 V, 50 Hz of nominal frequency, and stationary and/or transient disturbances.

Furthermore, as the noise is ever-present in PDN, the proposed methodology is also checked in noisy situations.

### 1) Training stage

In this step, 100 patterns of each different single and multiple-combined power quality disturbances are applied to linear and nonlinear loads. The following ten classes are tested:

- Harmonics (C0);
- Sags (C1);
- Swells (C2);
- Oscillatory transients (C3);
- Flicker voltage fluctuations (C4);
- Harmonics and sag (C5);
- Harmonics and swell (C6);
- Oscillatory transient and sag (C7);
- Oscillatory transient and swell (C8);
- Oscillatory transient and harmonics (C9);

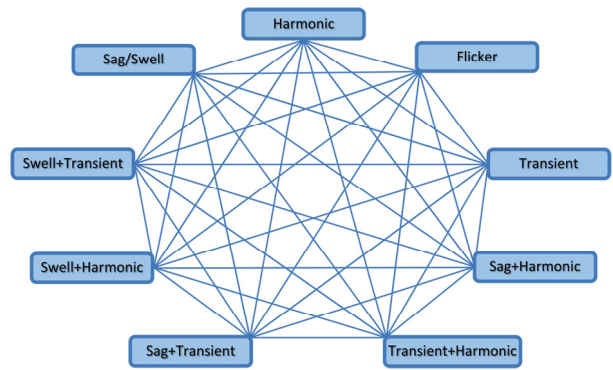

Fig. 4. One-versus-one training scheme for the selected disturbances classes.

Fig. 4 shows the different paired confrontation options for the selected disturbances classes in the one-versus-one training scheme. Here, nonstationary voltage variations corresponding to classes C1 and C2 are grouped only for graphic simplicity purposes. Fig. 6 depicts a 2-dimensional multiclass representation of each one of the classes used for the classification.

### 2) Noise conditions and Results

It is well stablished the adverse effect of noise over the performance of wavelet based event detection, the time localization and the classification schemes, due to the difficulties of separating noise and disturbances at higher frequencies bands [17], [18], [25]. In these cases, by using a suitable de-noising performance at the preprocessing stage, or at the featuring extraction stage, this limitation can be solved.

The nature of $ITD(t)$ index establishes an instantaneous ratio between the disturbance energy and the fundamental signal energy. In most of cases the event energy is higher than those corresponding to noise, consequently the $ITD(t)$ permits to detect the event and to measure its duration even in noisy signal corresponding to real-life situations. Fig. 5(a) depicts a disturbed signal generated in the programmable source and its $ITD(t)$ is shown in Fig. 5(c). A noisy version of this signal is measured in a load of the laboratory setup for testing experiments, and it is depicted in Fig. 5(b), where a high accuracy in the event duration calculation over $ITD(t)$ is possible yet as it can be seen in Fig. 5(d).

To test the manner in which the proposed method works under distinct noisy situations, different levels of Gaussian white noises with the signal to noise ratio (SNR) values from 30 to 50 dB have been considered. In these noisy conditions, the method works properly until the minimum value of 34 dB for the SNR. It corresponds to the limit case in which noise shows peak noise magnitude nearly 2% of the original input signal. Under this value, the compromised results correspond, first, to those events having a high sensibility to the $T_o$ parameter, and second, to those ones with a soft amplitude variation.

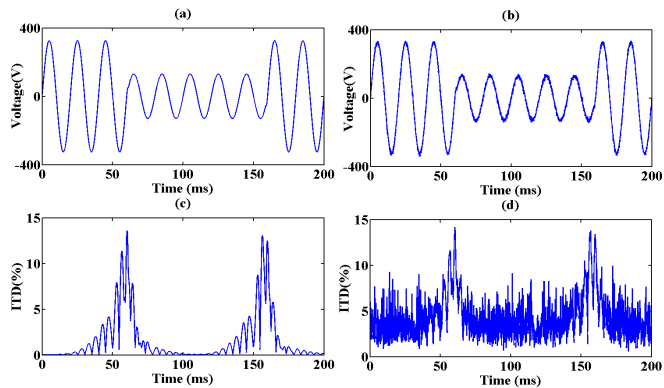

Fig. 5. a) Programed sag signal by the power supply. b) Sag with added noise signal measured in the load. c) ITD(t) of sag signal. d) ITD(t) of noisy sag signal.

In particular, because of noise, 10% of transient-sag events (C7) result in misclassified due to the similar features between C1-C7 classes. One of this case is shown in Fig. 5(a) and Fig. 5(b), where the input vectors ($k_1$, $k_2$) are (175.20, 2.58) and (175.46, 6.53) respectively. Observe that the increase of GDR parameter, $k_2$, in the second case delivers to confusion with the C7 zone class.

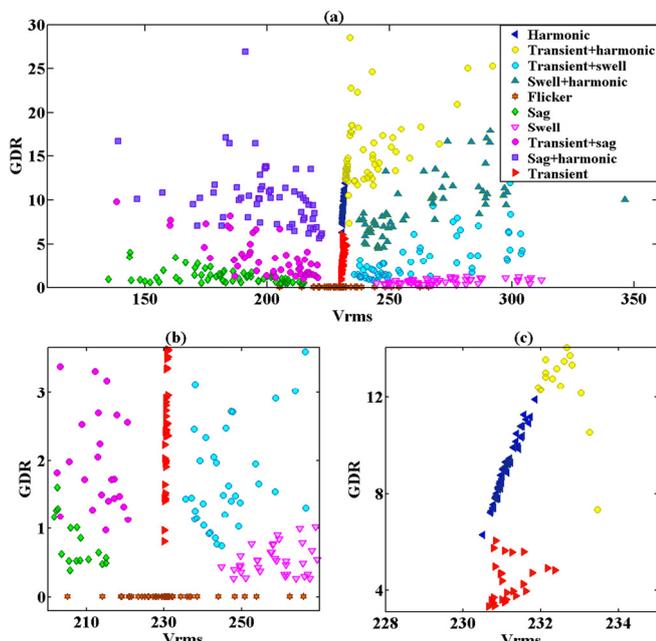

Fig. 6. a) 2D-feature vectors plot of the selected disturbances classes. b) Zoom over lower central area of Fig. 6(a). c) Zoom over upper central area.

Fig. 6(a) shows the plot of the 2D-feature vectors, which has to be classified by the SVM, and the different classes constrained to different zones. The details of two unclear classification regions can be seen in Fig. 6(b) and Fig. 6(c). Analogously, swell events with added noise in C2 class are sometimes confused with those of C8 class. On the other hand, the rest of the classes are accurately classified.

Table II summarizes the classification results based on a total of 1000 test event signals (100 per class), some of them with added noise and SNR upper or equal to 34 dB. Here, the successful classification results are tabulated at diagonal elements. On the contrary, the misclassification results are shown at non-diagonal elements. All these results reveal the robustness of the proposed method.

TABLE II
CLASSIFICATION RESULTS OF DISTURBED SIGNALS

| True class | Possible classes | | | | | | | | | | Accuracy (%) |
|---|---|---|---|---|---|---|---|---|---|---|---|
| | C0 | C1 | C2 | C3 | C4 | C5 | C6 | C7 | C8 | C9 | |
| C0 | 97 | 0 | 0 | 1 | 0 | 0 | 0 | 0 | 0 | 2 | 97 |
| C1 | 0 | 89 | 0 | 0 | 0 | 0 | 0 | 11 | 0 | 0 | 89 |
| C2 | 0 | 0 | 89 | 0 | 0 | 0 | 0 | 0 | 11 | 0 | 89 |
| C3 | 2 | 0 | 0 | 98 | 0 | 0 | 0 | 0 | 0 | 0 | 98 |
| C4 | 0 | 0 | 0 | 0 | 100 | 0 | 0 | 0 | 0 | 0 | 100 |
| C5 | 0 | 0 | 0 | 0 | 0 | 97 | 0 | 3 | 0 | 0 | 97 |
| C6 | 0 | 0 | 0 | 0 | 0 | 0 | 98 | 0 | 0 | 2 | 98 |
| C7 | 0 | 3 | 0 | 0 | 0 | 3 | 0 | 94 | 0 | 0 | 94 |
| C8 | 0 | 0 | 3 | 0 | 0 | 0 | 3 | 0 | 94 | 0 | 94 |
| C9 | 2 | 0 | 0 | 0 | 0 | 0 | 2 | 0 | 0 | 96 | 96 |
| Overall Success rate (%) | | | | | | | | | | | 94.2 |

## IV. CONCLUSION

An integral assessment of the electrical network PQ by means of the wavelet-based global disturbance ratio indicator (GDR) and the well-known RMS value of the signal is proposed. The GDR index is based on the instantaneous index ITD(t), which considers two PQ aspects: steady-state PQ relative to harmonic level, and the non-stationary PQ relative to oscillatory transients or sudden amplitude changes of the signal.

The possible DWT errors due to possible system frequency variation was performed as a pre-processing step. Ten complete periods of the signal under study are contained in the observation window using the IFC. Then, the number of stages of DWT is guaranteed to be a power of two.

The developed PQ System Classifier, based on wavelet techniques and SVM, is an effective device for detecting, monitoring and classifying stationary and nonstationary multi-event signals, even in three-phase systems. It is essentially a low-cost method of classification, and it evaluates on-line the signal quality by identifying single or multiple-combined disturbances in a PDN.

The low computational cost is due mainly to both, the choice of only six wavelet decomposition levels at the feature extraction stage, and the simplicity of the double-indices-based training process in the classification stage.

Therefore, a method like this, based on SVM and WMRA, gives computation simplicity, accuracy and shows a good compromise of speed, adaptability, and size of the time-frequency window. Moreover, a low-cost hardware prototyping based on the proposed method is possible and would add to the mentioned advantages, a great potential in the implementation of smart monitoring systems with on-line classification options. The results show that the proposed methodology is highly reliable and efficient.

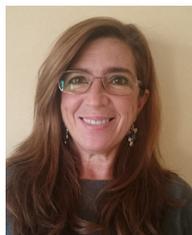
**Maria Dolores Borrás** received the B.Sc. degree in Physics Sciences, with specialization in electronics, from the University of Seville in 1990. She is working towards the Ph.D. degree in Electrical Engineering at the University of Seville.

Since 1992 she has been an Associate Professor in the Department of Electrical Engineering, University of Seville, Spain. She has participated in different projects related to Power Quality. She belongs to the Spanish Research Group "Invespot". Her research interests include digital signal processing, measurement of power quality in poly-phase systems, control of renewable energy systems, and smart-grid solutions.

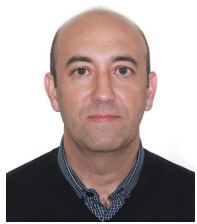
**Juan Carlos Bravo** received the PhD. degree from the University of Seville, Spain, in 2008.

He is currently an Associate Professor in the Department of Electrical Engineering, University of Seville, Spain. He is currently the head of the Spanish Research Group "Invespot" on Power Quality. His research interests include geometric algebra applied to power quality in poly-phase systems, harmonics and power systems assessment via time-frequency transforms analysis.

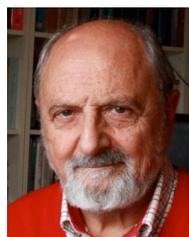
**Juan Carlos Montaño** (M'80, SM'00 and LSM'12) received the PhD degree in physics from the University of Seville, Spain, in 1972.

From 1973 to 1978 he was a Researcher at the Instituto de Automática Industrial (CSIC-Spanish Research Council), Madrid, Spain, working on analog signal processing, electrical measurements, and control of industrial processes. Since 1978, he has been responsible for various projects in connection with research in power theory of nonsinusoidal systems, reactive power control, and power quality at the CSIC. In 1989 he founded the Spanish mixed research group (CSIC-University of Seville) named "Invespot".  He is the author of many refereed publications in international journals, book chapters, and conference proceedings. He has also directed several investigation projects founded by public institutions and private industry. He has been responsible of a coordinated project, between the CSIC and the University of Seville, for the study of electrical fault prevision in the line transmission and the effects of signal disturbances on grid-connected renewable power plants. Recently, he is participating at the University of Seville, in a project related to wireless intelligent sensor networks for smart grids.